\documentclass[
 reprint,
superscriptaddress,
amsmath,amssymb,
aps,
pra,
]{revtex4-2}

\usepackage{graphicx}
\usepackage{dcolumn}
\usepackage{bm}
\usepackage{gensymb}
\usepackage{url}
\usepackage{multirow}

\usepackage{xcolor}

\begin{document}

\preprint{APS/123-QED}

\title{Suppressed-scattering windows for radiative cooling applications}

\author{José M. Pérez-Escudero}
\affiliation{%
 Department of Electrical, Electronic and Communications Engineering, Institute of Smart Cities (ISC), Public University of Navarre (UPNA), 31006 Pamplona, Spain
}
\author{Alicia E. Torres-García}
\affiliation{%
 Department of Electrical, Electronic and Communications Engineering, Institute of Smart Cities (ISC), Public University of Navarre (UPNA), 31006 Pamplona, Spain
}
\author{Carlos Lezaun}
\affiliation{%
 Department of Electrical, Electronic and Communications Engineering, Institute of Smart Cities (ISC), Public University of Navarre (UPNA), 31006 Pamplona, Spain
}

\author{Antonio Caggiano}
\affiliation{%
DICCA, Dept. of Civil, Chemical and Environmental Engineering, University of Genova, Via Montallegro 1, Genova 16145, Italy
}

\author{Ignacio Peralta}
\affiliation{%
Institut für Werkstoffe im Bauwesen, Technische Universität Darmstadt, Germany
}
\affiliation{%
Centro de Investigación de M\'etodos Computacionales (CIMEC), Universidad Nacional del Litoral (UNL)/ Consejo Nacional de Investigaciones Cient\'ificas y T\'ecnicas (CONICET), Predio CONICET ``Dr. Alberto Cassano", Colectora Ruta Nac. 168 km 0, Paraje El Pozo, 3000 Santa Fe, Argentina
}
\affiliation{%
Laboratorio de Flujometr\'ia (FLOW), Facultad Regional Santa Fe (FRSF), Universidad Tecnol\'ogica Nacional (UTN), Lavaise 610, 3000 Santa Fe, Argentina
}

\author{Jorge S. Dolado}
\affiliation{%
Centro de Física de Materiales CFM (CSIC-UPV/EHU), Donostia-San Sebastián, Spain
}
\affiliation{%
Donostia International Physics Center (DIPC), Donostia-San Sebastián, Spain
}

\author{Miguel Beruete}
\affiliation{%
 Department of Electrical, Electronic and Communications Engineering, Institute of Smart Cities (ISC), Public University of Navarre (UPNA), 31006 Pamplona, Spain
}

\author{Iñigo Liberal}
\thanks{Corresponding author: inigo.liberal@unavarra.es}%
\affiliation{%
 Department of Electrical, Electronic and Communications Engineering, Institute of Smart Cities (ISC), Public University of Navarre (UPNA), 31006 Pamplona, Spain
}%

\begin{abstract}
The scattering of light by resonant nanoparticles is a key process for enhancing the solar reflectance in daylight radiative cooling. Here, we investigate the impact of material dispersion on the scattering performance of popular nanoparticles for radiative cooling applications. We show that, due to material dispersion, nanoparticles with a qualitatively similar response at visible frequencies exhibit fundamentally different scattering properties at infrared frequencies. It is found that dispersive nanoparticles exhibit suppressed-scattering windows, allowing for selective thermal emission within an highly reflective sample. The existence of suppressed-scattering windows solely depends on material dispersion, and they appear pinned to the same wavelength even in random composite materials and periodic metasurfaces. Finally, we investigate calcium-silicate-hydrate (CSH), the main phase of concrete, as an example of a dispersive host, illustrating that the co-design of nanoparticles and host allows for tuning of the suppressed-scattering windows. Our results indicate that controlled nanoporosities would enable concrete with daylight passive radiative cooling capabilities.
\end{abstract}

\maketitle

\section{Introduction}

\begin{figure}[t]
\centering
\includegraphics[width = 0.35\textwidth]{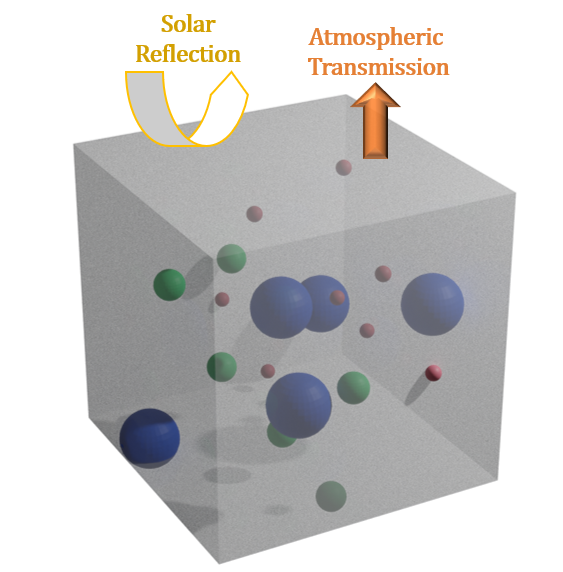}
\caption{Sketch of a radiative cooling composite material, consisting of a mixture of resonant nanoparticles reflecting solar light while efficiently radiating heat through the atmospheric window.}
\label{Sketch}
\end{figure}

Radiative cooling has emerged as a passive technology with the potential for a major reduction of energy consumption. It promises to cut the costs of cooling systems, alleviate the urban island effect, and combat global warming \cite{fan2022photonics,Hossain2016radiative,Zhao2019radiative,Yin2020terrestrial}. In essence, passive radiative cooling takes advantage of thermal emission to transmit energy through the atmospheric transparency window, using outer space as a cold sink. 
Achieving daytime and passive radiative cooling poses the additional technological challenge of minimizing sunlight absorbtion, while simultaneously maintaining a high infrared emissivity in the atmospheric transparency window (see Fig.\,\ref{Sketch}). 

However, this challenge can be met by nanophotonic design.
The seminal work by Raman et al. \cite{Raman2014passive} demonstrated that nanophotonic multilayered structures make passive daytime radiative cooling possible. Since then, multilayered \cite{Zeyghami2018review,Cunha2020multilayer,Kecebas2020,Mira2021,Mabchour2020daytime} and more advanced metamaterial structures \cite{Hossain2015metamaterial,Goyal2020recent} have been actively investigated. While nanophotonic engineering provides an excellent radiative cooling performace, the need to rely on complex nanofabrication processes and expensive materials limits its large-scale and low-cost deployment.
Composite materials provide an alternative approach towards daytime passive radiative cooling without the need of nanofabrication process and/or expensive materials. Instead, composite materials use random mixtures of nanoparticles, whose size distribution is engineered to support the resonances needed for efficient solar reflectance.
This concept has proven to be very successful, leading to cost-effective solutions for radiative cooling structures in the form of microsphere coatings and paints \cite{Atiganyanun2018,li2021ultrawhite,Cheng2019effect,Huang2017}, polymers \cite{Zhai2017scalable,Mandal2018hierarchically,li2021scalable,Gao2021cooling}, biomimetics \cite{Shi2018nanostructured,shi2015keeping}, textiles \cite{Zeng2021hierarchical,Zhu2021subambient} and structural materials like wood \cite{Li2019radiative}.
 
The nanophotonic design of composite materials for radiative cooling has predominantly focused on the use of Mie scattering resonances to enhance solar reflectance. However, scattering theory is a large and mature field, and many of the ideas developed through years of research can be applied to the design of radiative cooling systems. 
Indeed, the interplay between material dispersion and geometry enables a wealth of scattering phenomena including resonantly enhanced scattering and absorption processes \cite{Ruan2010superscattering,krasnok2019anomalies,liberal2014upper}, scattering cancellation and invisibility \cite{Alu2005achieving}, superdirectivity \cite{Arslanagic2018highly,Ziolkowski2017using}, cloaked (minimum scattering) sensors \cite{Alu2009cloaking,liberal2013analytical}, superbackscattering \cite{liberal2015superbackscattering,powell2021multiband,liberal2015superbackscatteringarray}, frequency comb scattering \cite{Monticone2013multilayered}, Fano resonances \cite{Limonov2017fano}, nonradiating epsilon-near-zero \cite{liberal2016nonradiating,Monticone2014embedded,silveirinha2014trapping} and anapole \cite{Miroshnichenko2015nonradiating} modes, as well as the enhancement \cite{Bharadwaj2009optical,lodahl2015interfacing} and inhibition \cite{Yablonovitch1987inhibited,Arslanagic2013jamming,gong2022radiative} of the radiation by quantum emitters.
The large body of work on scattering theory suggests that the interplay between dispersion and geometry should be carefully examined when choosing a material system for radiative cooling applications. Following this motivation, here we analyze the impact of material dispersion on the scattering performance of nanoparticles that are commonly employed to enhance the solar reflectance of radiative cooling systems. As we will show, several of them exhibit ``suppressed-scattering windows" (i.e., the suppression of scattering for a wide range of nanoparticle radii) overlapping with the atmospheric transparency window, providing a key insight in the design of composite materials for daytime passive radiative cooling.  

\begin{figure*}[t]
\centering
\includegraphics[width = 0.95\textwidth]{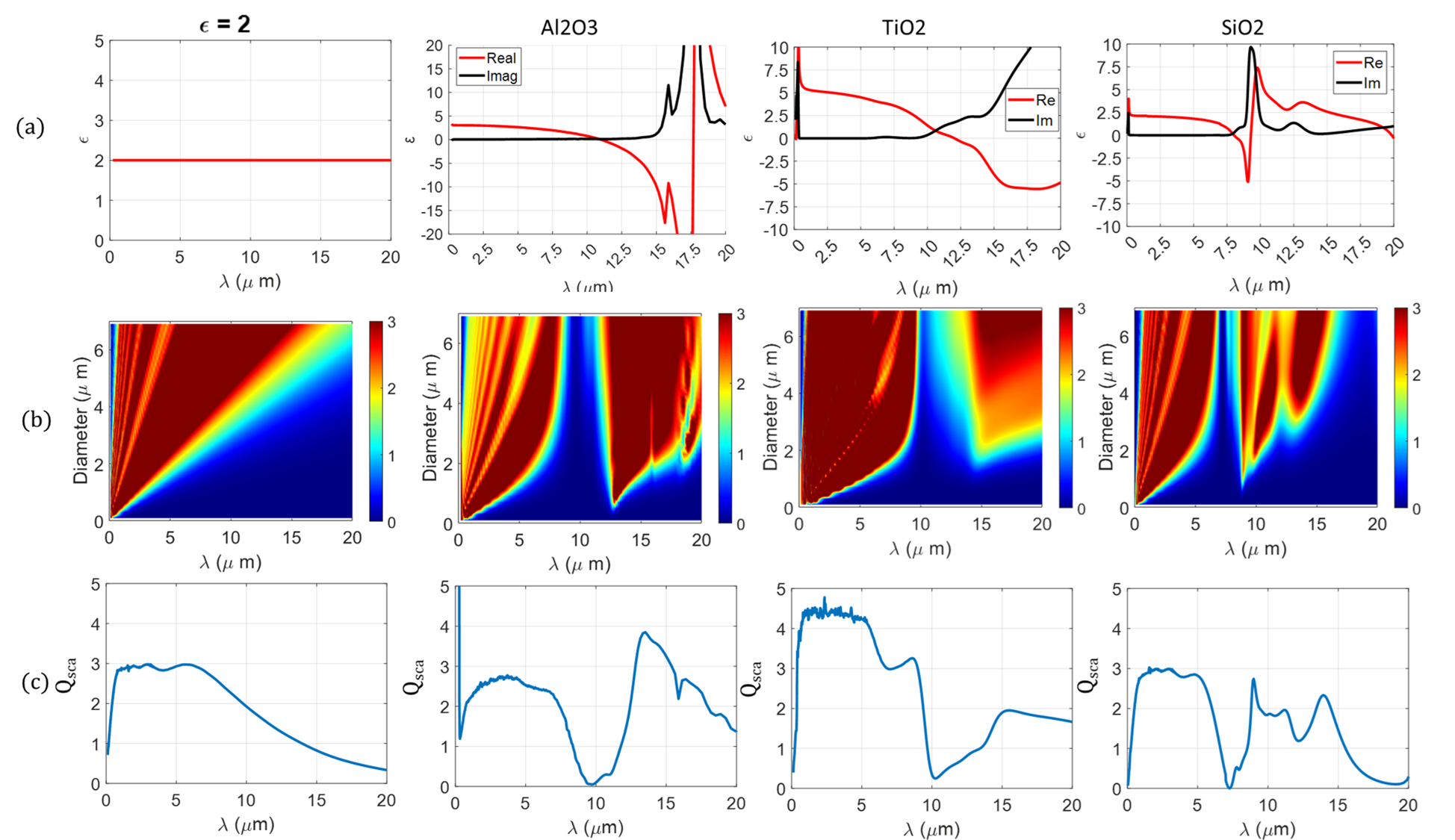}
\caption{Scattering efficiency calculations for nondispersive dielectric, alumina ($Al_2O_3$), titanium dioxide ($TiO_2$) and silica ($SiO_2$) nanoparticles: (a) Frequency dispersion of the complex permittivity. Real part shown in solid red line and imaginary part in solid black line. (b) Scattering efficiency for the nanoparticles embedded in air ($\varepsilon_h=1$). (c) Average value of the scattering efficiency for nanoparticle radii ranging from $0$ to $7\,\mu m$.}
\label{materialsair}
\end{figure*}

\section{Scattering from dispersive nanoparticles \label{sec:Individual}}

We start by analyzing the individual scattering performance of popular nanoparticles used in radiative cooling applications. As case studies, we consider alumina ($Al_2O_3$), titanium dioxide ($TiO_2$) and silica ($SiO_2$) nanoparticles, which are commonly employed for increasing the solar reflectance \cite{Wenxin2020,Liu2019}. For comparison, we include in the analysis nanoparticles made of an ideally nondispersive material, with dielectric constant $\varepsilon_{ND}=2$ at all frequencies. Fig.~\ref{materialsair}(a) depicts the frequency dispersion of their permittivities following tabulated data (see Methods: Material parameters). It can be concluded from the figure that, at visible frequencies, all materials behave approximately as nondispersive dielectrics with negligible loss. Therefore, their frequency response in this range is qualitatively similar, the main difference being their dielectric constant value. For example,  
$\varepsilon_{ND}=2$,
$\varepsilon_{Al_2O_3}\simeq 3.05$,
$\varepsilon_{TiO_2}=6.17$,
and $\varepsilon_{SiO_2}=2.10$,
at $\lambda=1\,\mu m$,
with $TiO_2$ exhibiting a significantly larger value.
Therefore, all studied materials are expected to support dielectric (Mie-scattering) resonances in the visible, enhancing the solar reflectante. It justifies their choice as popular materials for radiative cooling applications.  By contrast, their material response at infrared frequencies presents a more involved resonant response, raising qualitative differences between the materials. For example, while $Al_2O_3$ is characterized by relatively sharp material resonances at frequencies below the atmospheric window, $TiO_2$ exhibits a broad resonance band at those wavelengths. As for $SiO_2$, it features a sharp resonant peak within the atmospheric window. Therefore, the scattering performance of these materials is expected to be very different at infrared frequencies, highlighting the need to take into account material dispersion for their implementation within radiative cooling systems.

Fig.\,\ref{materialsair}(b) depicts the scattering efficiency of nanoparticles made of these materials as a function of wavelength and nanoparticle radius, computed by using Mie-scattering theory (see Methods: Calculation of the scattering efficiency). For the nondispersive dielectric material, the scattering efficiency exhibits multiple resonant peaks with a linear dispersion in wavelength and nanoparticle size. Within the $0.2-2.0 \mu m$ frequency band, a qualitatively similar behavior is obtained for $TiO_2$, $Al_2O_3$ and $SiO_2$ nanoparticles. Thus, it is confirmed that in this frequency range these materials effectively behave as nondispersive dielectrics. Minor differences arise due to the higher value of the $TiO_2$ dielectric constant, yielding more densely packed resonances, and with higher peak values. 

By contrast, the impact of material dispersion in the scattering performance is more crucial at infrared frequencies. Due to material dispersion, the nanoparticle resonances exhibit a highly nonlinear dispersion. In addition, the scattering efficiency maps reveal several frequency bands where the scattering efficiency is suppressed for all considered nanoparticle radii. Comparison with Fig.\,\ref{materialsair}(a) reveals that these minima correspond to frequencies $\omega_1$ where the real part of the permittivity approximately equals one, i.e., $\varepsilon\left(\omega_1\right)\simeq 1$. More generally, when the dielectric permittivity of the nanoparticle matches that of its host matrix, the scattered field becomes identically zero, irrespectively of the size and shape of the nanoparticle. Therefore, the scattering is minimized for any nanoparticle radius, forming a ``suppressed scattering window''. While the origin of such suppressed-scattering window is conceptually simple, we believe that their important technological implications, with realistic materials, and for radiative cooling applications, has not been fully explored. As we will show, it should critically condition the choice of the nanoparticle material. 

Remarkably, the suppressed-scattering windows of known nanoparticles such as $TiO_2$ and $Al_2O_3$ overlap with the atmospheric transparency window. Therefore, nanoparticle systems based on $TiO_2$ and $Al_2O_3$ will exhibit a transparency band where thermal emission is needed the most. This property facilitates the design of selective thermal emitters for radiative cooling applications. For example, if a material is a good thermal emitter but has poor solar reflectance, one could add a layer of these nanoparticles on top of it, increasing its solar reflectance while guaranteeing that the suppressed scattering window allows for thermal emission within that frequency band. Alternatively, one could add nanoparticles within a good thermal emitter. Selecting nanoparticles whose permittivity match that of the host in the atmospheric window one could again increase the solar reflectance, while minimizing the impact on the thermal emission performance. 
On the contrary, $SiO_2$ nanoparticles exhibit strong resonances within the atmospheric window, which will result in undesired reflectivity. At the same time, nanoparticle resonances enhance the absorption of poor thermal emitters, particularly if it has a small volume. For this reason, $SiO_2$ might be benefitial for thin and electrically small radiative coolers, or for cases when the matrix itself is a poor thermal emitter.

These conclusions highlight that the dispersion profile of resonant nanoparticles must be carefully taken into account, as different profiles lead to different strategies in the design of radiative coolers. 
To further illustrate this point, Fig.~\ref{materialsair}(c) depicts the average scattering efficiency for nanoparticle radii ranging from 0 to 7 $\mu$m. The results illustrate the presence of the suppressed-scattering window for any nanoparticle radii, and the different scattering performance of $TiO_2$, $Al_2O_3$ and $SiO_2$ at infrared frequencies.

\section{Reflection from random mixtures of nanoparticles \label{sec:Composites}}

Radiative cooling systems augmented with resonant nanoparticles typically consist of complex composite materials, containing distributions of nanoparticles of different sizes in a random arragement. Moreover, the size of resonant nanoparticles is comparable to, or larger than, the wavelength of operation, posing additional difficulties on the theoretical modelling. Therefore, providing an accurate estimation of the radiative cooling performance of a composite material containing densely packed resonant nanoparticles is a cumbersome task. An accurate prediction of the power ratios and cooling rates of a composite material would require from large-scale and stochastic full-wave numerical simulations. However, an appproximate estimation can be obtained with the use of effective medium theories (EMTs) \cite{Sihvola1999}, where extensions to conventional EMTs allow for the analysis of composites containing large and resonant particles \cite{Bijarniya2020review,niklasson1989radiative,Bohren1986applicability}. In this section, we use these approximate methods to conceptually validate the extension of the conclusions drawn in the previous section to random mixtures of nanoparticles.

\begin{figure*}[t]
\centering
\includegraphics[width = 0.95\textwidth]{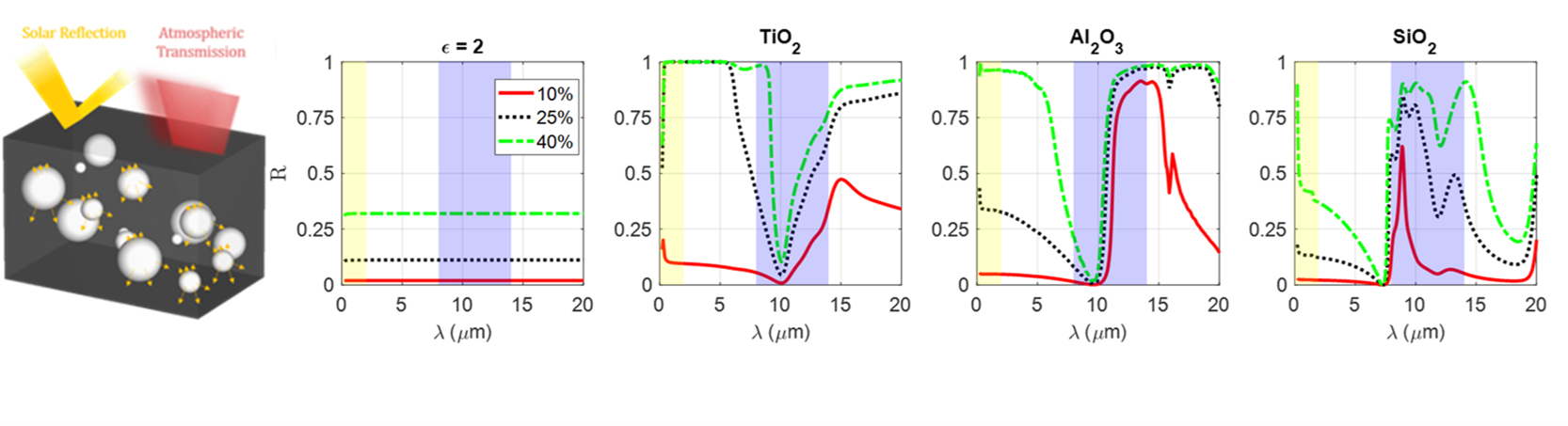}
\caption{Calculated reflectivity for a composite material consisting of an air host medium containing a random distribution of nanoparticles made of (a) a nondispersive dielectric material ($\varepsilon_p = 2$), (b) titanium dioxide, (c) alumina and (d) silica, for different filling factors $f=10\%,\,25\%\,{\rm and}\,40\%$. The nanoparticles size distribution ranges from 50 nm to 300 nm radius, with a step of 50 nm.}
\label{Rtotal}
\end{figure*}

Fig.\,\ref{Rtotal} depicts the reflectivity $R =  \left |(1-\sqrt{\varepsilon_e / \mu_e})/(1+\sqrt{\varepsilon_e / \mu_e })\right|^2$ 
from composites containing a distribution of resonant nanoparticles, modeled with effective permittivity ($\varepsilon_e$) and permeability ($\mu_e$), calculated via generalized Maxwell-Garnett theories (see Methods: Calculation of the reflectivity from composite materials).
Following the analysis of the scattering of individual nanoparticles, we consider mixtures of nanoparticles made of the same four materials, all immersed in a transparent, unit-permittivity host. For all four classes of mixtures, we consider a distribution of nanoparticle sizes from 50nm to 300nm with a step of 50nm. In all four cases, we study mixtures with three different filling factors of 10\%, 25\% and 40\%.

We note that the EMT takes into account the interaction between the nanoparticles, and the results might in principle deviate from those of the analysis of individual nanoparticles, particularly for large filling factors. 
However, the results reported in Fig.\,\ref{Rtotal} qualitatively match the conclusions drawn in Section\,\ref{sec:Individual}. Specifically, it is found that the reflectivity spectra for the nondispersive nanoparticles remain flat, showing no particular resonances. On the other hand, dispersive nanoparticle materials, like $TiO_2$, $Al_2O_3$ and $SiO_2$, feature a clear reflectivity dip, whose wavelength exactly corresponds to the suppressed-scattering window observed at the individual nanoparticle level (see Fig.\,\ref{materialsair}). In fact, the reflectivity dip remains fixed at the same wavelength even for large filling factors, for which the reflectivity in the rest of the spectrum is high. Hence, it is confirmed that the interaction between the nanoparticles does not affect the existence and location of the reflectivity minima. 

It is found that composites made of $TiO_2$ and $Al_2O_3$ act as very selective materials, exhibiting a high reflectivy at wavelengths shorter and longer than that of the suppressed-scattering window.
Again, material dispersion is key for understanding the overall response of the composite. At wavelengths shorter than the suppressed-scattering window, high reflectivity is obtained via Mie-scattering resonances. At longer wavelengths, high reflectivity is associated with the negative value of the nanoparticles permittivity (see Fig.\,\ref{materialsair}(a)), effectively acting as a nonresonant and homogeneous mirror. Thus, our results demonstrate that by exploiting material dispersion it is possible to design highly-reflective composite materials, which nevertheless remain transparent in a selective window.  

\section{Extension to metasurfaces and different nanoparticle shapes \label{sec:Metasurfaces}}

\begin{figure*}[t]
\centering
\includegraphics[width = 0.99\textwidth]{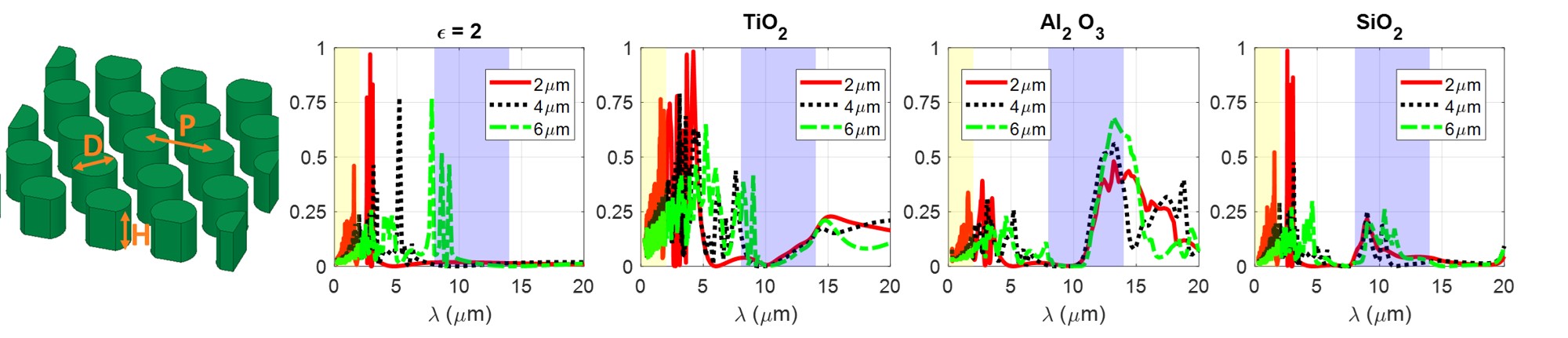}
\caption{Calculated reflectivity for a metasurface consisting of an hexagonal array of truncated cylinders made of (a) a nondispersive dielectric material ($\varepsilon_p = 2$), (b) titanium dioxide, (c) alumina and (d) silica, for diffent cylinder diameters.}
\label{GDCALC}
\end{figure*}

Since suppressed-scattering windows are solely induced by the material properties of the nanoparticles, their existence should be guaranteed independently of the geometry of the system. Therefore, suppressed-scattering windows should be observed irrespectively of the shape (cylindrical, elliptical or arbitrary) and/or the arrangement (metasurfaces, photonic crystals, metamaterials) of the nanoparticles. 
To illustrate this point, Fig.\,\ref{GDCALC} depicts the reflectivity of a metasurface consisting of a hexagonal array of small truncated cylinders. The aspect ratio of the cylinders, $AR$ = height ($H$) / diameter ($D$), is set to $AR=1$, and period of the arrays is set to $P=3/2\cdot D$. In order to assess the geometry-independence of suppressed-scattering windows, the diameter of the cylinders is varied from $D=2\,\mu$m to $D=6\,\mu$m. The results were obtained with a full-wave numerical solver (see Methods: Numerical Calculations).
The calculated reflectivities are characterized by a discrete spectrum with multiple resonant peaks, consistent with the periodic nature of the geometry. In general, the positioning of the resonances critically depends on the size of the truncated cylinders. 

However, it is observed that, for the arrays of $TiO_2$, $Al_2O_3$ and $SiO_2$ dispersive cylinders, the reflectivity is consistently suppressed near the same wavelengths than the minima observed at the individual particle level (see Fig.\,\ref{materialsair}), independently of the geometrical configuration of the metasurface. Therefore, the numerical simulations demonstrate that our results can be extended to particles of arbitrary shape, as well as to regular arrangements. Thus,  suppressed-scattering windows open the possibility of designing metasurfaces with strong resonance effects at specific wavelengths, while ensuring the existence of a fixed transparency band. We expect that many of the applications of Mie-resonant metaphotonics / Mie-tronics \cite{kuznetsov2016,koshelev2020dielectric} might benefit from this simple design principle. Materials like $TiO_2$ and $Al_2O_3$, whose suppressed-scattering windows overlap the atmospheric transparency window, seem particularly suitable not only for radiative cooling, but for any radiative thermal engineering applications. 
 
\section{From dispersive nanoparticles to dispersive hosts \label{sec:Concrete}}

\begin{figure}[t]
\centering
\includegraphics[width = 0.35\textwidth]{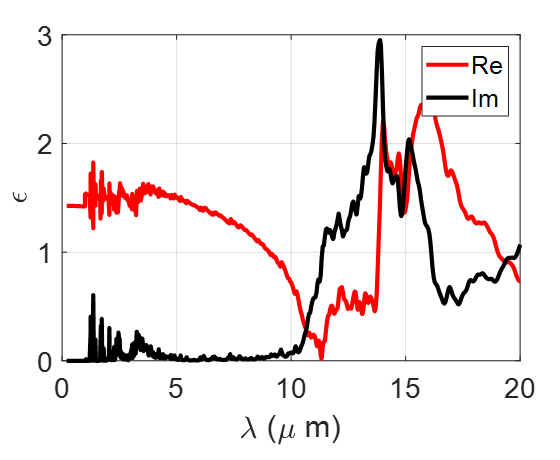}
\caption{Frequency dispersion of the permittivity of Calcium-Silicate-Hydrate (CSH).}
\label{fig:Permittivity_CSH}
\end{figure}

Up to this point, our analysis has focused on dispersive nanoparticles embedded in a nondispersive host. However, it is expected that a co-design of nanoparticle and host material dispersion will be needed in many practical cases. Here we investigate the scattering performance of nanoparticles embedded in a dispersive calcium-silicate-hydrate (CSH) host. CSH is of great technological interest as it is the main phase of concrete, thus with the potential to critically impact the practical deployment of radiative cooling systems. The permittivity of CSH is depicted in Fig.\,\ref{fig:Permittivity_CSH} as computed from atomistic simulations (see Methods: Material parameters). It can be concluded from the figure that CSH behaves as a nearly-nondispersive dielectric at visible frequencies, while it features several material resonances within the atmospheric window. Therefore, CSH is a good thermal emitter, but its reduced solar reflectance limits its applicability for daylight operation. 

\begin{figure*}[t]
\centering
\includegraphics[width = 0.98\textwidth]{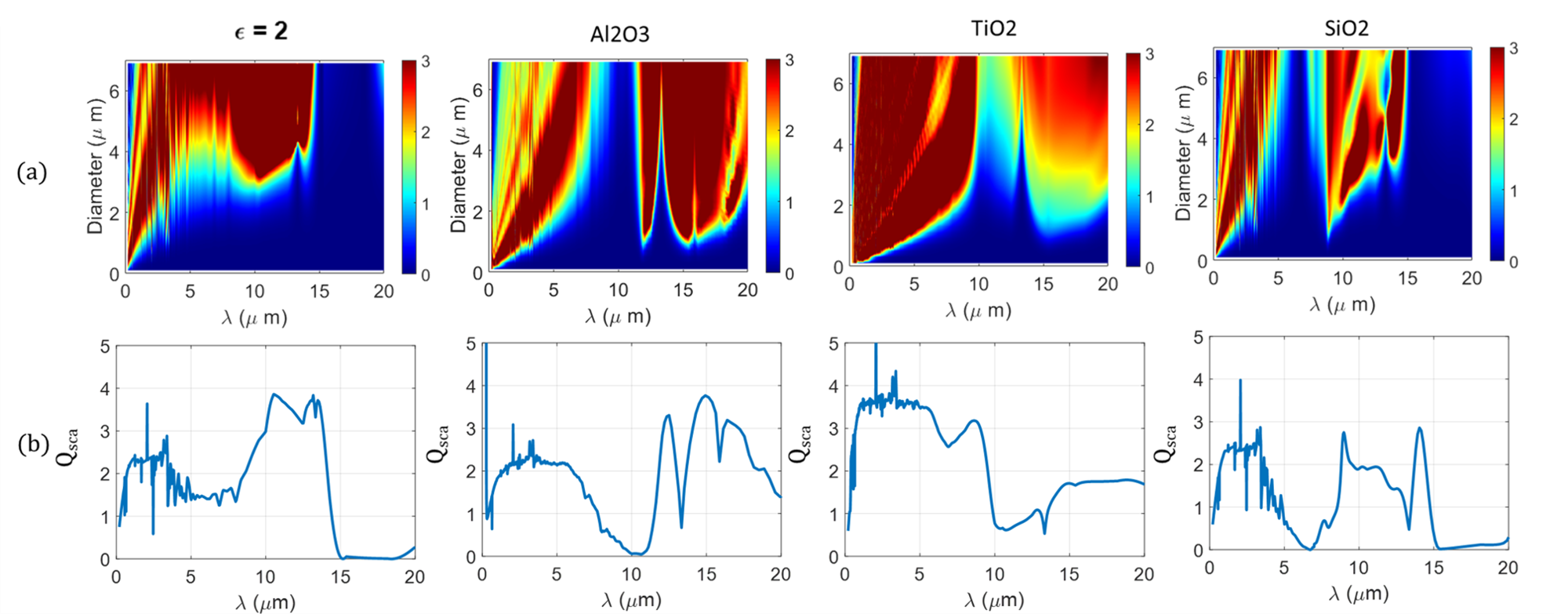}
\caption{(a) Scattering efficiency $Q_{sca}$ calculations for nondispersive dielectric ($\varepsilon_p=2$), alumina ($Al_2O_3$), titanium dioxide ($TiO_2$) and silica ($SiO_2$) nanoparticles, embedded in a calcium-silicate-hydrate (CSH) host. (c) Average value of the scattering efficiency for nanoparticle radii from $0$ to $7\mu m$.}
\label{materialshost}
\end{figure*}

However, the solar reflectivity of CSH can be enhanced by adding resonant nanoparticles and/or by controlling its internal porosity. Fig.~\ref{materialshost} shows the scattering efficiency spectra as a function of the nanoparticle radii for nondispersive ($\varepsilon_p=2$), $Al_2O_3$, $TiO_2$ and $SiO_2$ nanoparticles immersed in a CSH host. As expected, the dispersion of the host material impacts the scattering efficiency, and, in particular, the existence of  suppressed-scattering windows. For example, nondispersive nanoparticles do not exhibit any suppressed-scattering window when immersed in air (see Fig.\,\ref{materialsair}), but they are shown to exhibit a  suppressed-scattering window at long wavelengths when immersed in CSH. Furthermore, the frequency positioning and the depth of the suppressed-scattering windows of $Al_2O_3$, $TiO_2$ and $SiO_2$ nanoparticles are affected by the dispersion of CSH. Specifically, the suppressed-scattering window of $SiO_2$ is shifted to shorter wavelenghts, and a second window appears at longer wavelenghts. Similarly, the suppressed-scattering window of $TiO_2$ is shifted, and features a smaller depth. On the contrary, the suppressed-scattering window of $Al_2O_3$ retains a considerable depth for all the studied radii, suggesting that it might be the best candidate to enhance the radiative cooling performance of CSH. In general, our results highlight the need of codesigning the material dispersion of nanoparticles and host in order to achieve the best performance.

\begin{figure}[t]
\centering
\includegraphics[width = 0.35\textwidth]{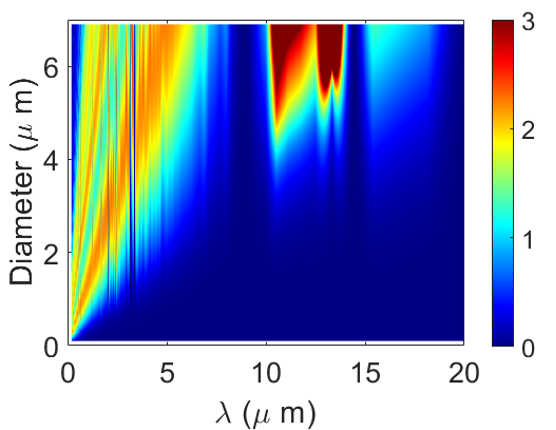}
\caption{(a) Scattering efficiency $Q_{sca}$ calculations for porosity bubbles ($\varepsilon_p=1$) embedded in a calcium-silicate-hydrate (CSH) host.}
\label{fig:Qscat_CSH}
\end{figure}

Finally, we investigate the scattering performance of of air ($\varepsilon_p=1$) bubbles of different size within a host CSH material, modeling nanoporous in CSH. The scattering efficiency, presented in Fig.\,\ref{fig:Qscat_CSH}, reveals that the porosity introduces Mie resonances in the visible, with the potential to enhance the solar reflectivity. At the same time, CSH porosity is found to exhibit a suppressed-scattering windows around 9\,$\mu$m, allowing for selective thermal emission in the atmospheric window. Therefore, CSH porosity has the key ingredients to boost radiative cooling performance, and controlled nanoporosity might open a pathway for efficient concrete-based daylight radiative cooling. 

Nanometer-scale length of cement-based materials is dominated by the so-called colloidal-scale porosity within CSH gel phases \cite{Jennings2007multi,zhang2019effects} (i.e., arising from the colloidal particle dispersion in porous media \cite{johnson1996colloid}). It is known that nanoporosity can have huge effects on the overall thermo-physical and mechanical properties of cement systems \cite{jennings2008characterization,Allen2007composition,Azimi2020synergistic}. Our results demonstrate that nanoporosity will also have a profound impact on their radiative cooling performance. Multiple studies have been proposed to tune the inner pore structures and to sharp the performance of cement-based materials by employing supplementary materials such as reactive slags or by products like clay, metakaolin, silica fume, fly ash, graphene oxides, colloidal nano-silica, carbon nanofibers and nanotubes, etc. \cite{Barbhuiya2017nanoscaled,Jafarbeglou2015nanoscience,Horszczaruk2015nanocomposite,Luo2021nanoindentation}. Our results suggest that the impact of such tuning on the radiative cooling performance should be addressed, simultaneously optimizing optical material dispersion and scattering performance, and the thermal and chemical properties of cement systems.

\section{Conclusions \label{sec:Conclusions}}

We studied the scattering performance of popular nanoparticles employed to enhance the solar reflectance for radiative cooling applications, including ideally nondispersive nanoparticles, and nanoparticles made of popular materials in radiative cooling applications. All the studied nanoparticles approximately behave as nondisperive dielectrics in the visible, allowing for an improvement of the solar reflectance via Mie resonances. However, their response at infrared frequencies is found to be qualitatively different, highlighting the need to take into  account the frequency dispersion. We found that alumina ($Al_2O_3$), and titanium dioxide ($TiO_2$) nanoparticles exhibit  suppressed-scattering windows overlapping with the atmospheric window, while silica ($SiO_2$) nanoparticles exhibit strong resonances in the same frequency range. In turn, those differences lead to different design strategies for radiative coolers. It was shown that suppressed-scattering windows result in robust and selective transmissive bands within an highly reflecting window. Our results highlight a key insight in the design of selective thermal emitters for radiative cooling applications, and it can be applied in a variety of materials and technological platforms. 

We used a generalized Maxwell-Garnett theory to demonstrate that our conclusions extend to random mixtures of nanoparticles. We expect that the intuitions obtained from scattering theory will provide helpful design guidelines for composite materials. We also used numerical simulations to demonstrate that the concept of a  suppressed-scattering windows can be applied to metasurfaces and other regular structures. The possibility of fixing a transparency window while independently tunning resonant effects at other wavelenghts might find applications in Mie-resonant metaphotonics. 

Finally, we analyzed the scattering performance of the porosity (air bubbles) of calcium-silicate-hydrate (CSH), acting as a dispersive host. Our analysis reveals that the porosity of CSH gathers the key ingredients for daylight passive radiative cooling operation. Since CSH is the main phase of concrete, our results motivate further research on the control of nanoporosity of concrete, with the potential to enable a mass depoloyment of daylight radiative cooling applications. 

\section*{Acknowledgements}

The authors acknowledge finantial support from European Union’s Horizon 2020 research and innovation programme under grant agreement 964450, MIRACLE.

\section*{Materials and methods}

\subsection{Material parameters}

The frequency dependence of the permittivity of titanium dioxide ($TiO_2$), alumina ($Al_2O_3$) and silicon dioxide ($SiO_2$) were obtained from tabulated data \cite{Querry1985optical,Malitson1965interspecimen,Tan1998determination,Kitamura2007optical,Devore1951refractive,Siefke2016materials}. For the nondispersive dielectric, $\varepsilon_p=2$ was assumed at all frequencies. The dispersive permittivity of calcium-silicate-hydrate (CSH) was obtained from atomistic simulations following \cite{Dolado2020}.

\subsection{Calculation of the scattering efficiency}

Scattering efficiencies were computed by using Mie theory \cite{Book_Bohren}. The scattering efficiency is defined as the scattering cross section normalized to the area of the particle

\begin{equation}
Q_{sca} = \frac{C_{sca}}{\pi a^2}
\label{Qcsa}
\end{equation}

\noindent where $a$ is the nanoparticle radius, and $C_{\rm csa}$ is the scattering cross-section 

\begin{equation}
C_{sca} = \frac{2\pi}{k^2}\sum_{n=1}^\infty (2n+1)(|a_n|^2+|b_n|^2)
\label{Csa}
\end{equation}

\noindent with $k=\omega/c\sqrt{\varepsilon_h}$ being the propagation constant in a host with permittivity $\varepsilon_h$. $a_n$ and $b_n$ are the scattering coefficients 

\begin{equation}
a_n = \frac{m \Psi_n(mx) \Psi'_n(x)-\Psi(x)\Psi'(mx)}{m \Psi_n(mx)\xi'_n(x)-\xi_n(x)\Psi'_n(mx)}
\label{coefa}
\end{equation}

\begin{equation}
b_n = \frac{\Psi_n(mx) \Psi'_n(x)-m\Psi(x)\Psi'(mx)}{\Psi_n(mx)\xi'_n(x)-m\xi_n(x)\Psi'_n(mx)}
\label{coefb}
\end{equation}

\noindent where $m=\sqrt{\varepsilon_p/\varepsilon_h}$ is the refractive index ratio, $x=ka$ is the size parameter, $\Psi_n(x)$ is the spherical Bessel function of the first kind and order $n$, $\xi_n(x)$ is the spherical Hankel function of the first kind and order $n$, and $f'_n(x)=df_n(x)/dx$, $f=\Psi,\xi$.

\subsection{Calculation of the reflectivity from composite materials}

The reflectivity from a composite material with effective permittivity $\varepsilon_e$ and effective permeability $\mu_e$ is given by \cite{Book_BalanisAEE}

\begin{equation}
R =  \left | \frac{1-\sqrt{\varepsilon_e / \mu_e}}{1+\sqrt{\varepsilon_e / \mu_e }} \right|^2
\label{R}
\end{equation}

Following generalized Maxwell-Garnett rules, $\varepsilon_e$ and $\mu_e$ can be directly linked to the scattering properties of individual nanoparticles and the filling factor $f$ as follows \cite{niklasson1989radiative,Bijarniya2020review} 

\begin{equation}
\epsilon_e = \varepsilon_h \frac{1+2\gamma /3}{1-\gamma /3}
\label{eeff}
\end{equation}

\begin{equation}
\mu_e = \frac{1+2\delta /3}{1-\delta /3}
\label{ueff}
\end{equation}

\noindent with

\begin{equation}
\gamma =  j\frac{3f}{2(ka)^3}\,\,(S_1(0) + S_1(\pi))
\label{ee}
\end{equation}

\begin{equation}
\delta = j\frac{3f}{2(ka)^3}\,\,(S_1(0) - S_1(\pi))
\label{ue}
\end{equation}

\noindent where $S_1(\theta)$ is angle-dependent scattering amplitude

\begin{equation}
S_1(\theta) = \sum_{n=1}^\infty \frac{2n+1}{n(n+1)}\left[a_n\frac{P_n^1(\cos\theta)}{sin\theta}+b_n\frac{\partial}{\partial\theta}P_n^1(\cos\theta)\right]
\label{S}
\end{equation}

\noindent where $P_n^1$ is the associate Legendre polynomial of order $n$ and degree $l=1$. 

\subsection{Numerical simulations}

Reflection from a metasurface, consisting of a regular array of truncated cylinders, was numerically obtained by using the Grating Diffraction Calculator (GD-Calc) MATLAB-based software for diffraction grating simulation using rigorous coupled-wave (RCW) theory \cite{GDCalc,GDCalc2}.

\bibliography{library}

\end{document}